# Energy Efficient Placement of ML-Based Services in IoT Networks

Mohammed M. Alenazi, Barzan A. Yosuf, Sanaa H. Mohamed, *Member, IEEE*, Taisir E. H. El-Gorashi, and Jaafar M. H. Elmirghani, *Fellow, IEEE*
School of Electrical Engineering
University of Leeds, Leeds, United Kingdom
{elmmal, b.a.yosuf, s.h.h.mohamed, t.e.h.elgorashi, j.m.h.elmirghani}@leeds.ac.uk

*Abstract*—The Internet of Things (IoT) is gaining momentum in its quest to bridge the gap between the physical and the digital world. The main goal of the IoT is the creation of smart environments and self-aware things that help to facilitate a variety of services such as smart transport, climate monitoring, e-health, etc. Huge volumes of data are expected to be collected by the connected sensors/things, which in traditional cases are processed centrally by large data centers in the core network that will inevitably lead to excessive transportation power consumption as well as added latency overheads. Instead, fog computing has been proposed by researchers from industry and academia to extend the capability of the cloud right to the point where the data is collected at the sensing layer. This way, primitive tasks that can be hosted in IoT sensors do not need to be sent all the way to the cloud for processing. In this paper we propose energy efficient embedding of machine learning (ML) models over a cloud-fog network using a Mixed Integer Linear Programming (MILP) optimization model. We exploit virtualization in our framework to provide service abstraction of Deep Neural Networks (DNN) layers that can be composed into a set of VMs interconnected by virtual links. We constrain the number of VMs that can be processed at the IoT layer and study the impact on the performance of the cloud fog approach.

## I. INTRODUCTION

Machine learning (ML) models have recently changed the basis for most modern-day applications that include object detection, natural language processing, self-driving cars, to name a few. As new technologies emerge, the adoption of ML based models widens. Studies forecast that billions of devices are now connected to the Internet and this figure is anticipated to rise further in the future[1]. Internet of Things (IoT) on its own is projected to generate five quintillions of data everyday day whilst on the other hand driverless cars are reported to generate up to 4TB worth of data every single hour of driving per day [2]. Thus, such an amount of data collected makes ML and Deep Neural Networks (DNNs) particularly attractive for deployment in the edge of the network. Traditionally, the collected data is transported from the edge all the way to the centralized cloud data center at the core network. However, the centralized processing approach introduces several challenges that include privacy, increased network power consumption due to the number of hops and unacceptable latency overheads [3], [4]. To address the aforementioned challenges, fog computing has been proposed in the literature as a decentralized processing paradigm by exploiting the resources available across the IoT-Cloud continuum. In most cases, the aggregate resources available in the IoT-cloud continuum are overlooked in favor of centralized processing using the cloud data center [5]. Processing, networking, and storage are examples of such resources that can be used to relieve the cloud. Fog allows for cloud-based services to be brought closer to the data source, facilitating effective and fast processing [6]. A fog node can be any device with CPU and networking capabilities [7]. Notwithstanding the benefits of fog computing, there still exists a number of obstacles that need to be addressed before it can reach its full potential. These include Interoperability, fog networking, orchestration and resource provisioning, and computation offloading, which are just some of the issues. We focus on the energy efficient placement of interconnected virtual machines (VMs) that are used to abstract DNN layers [3]. In this work, we extend our previous contribution in [8] by 1) studying the impact of a range of the idle power proportion ratio (IPPR) that is attributed to our application for highly shared networking equipment in the access, metro and core networks. This can also represent the growth of the given application segment considered in the future provided that the idle power consumption is proportional to the size of the application and 2) paying particular attention to the impact of the constraint imposed on the processing capability at the IoT layer such that each IoT node can only process a limited number of VMs. This is a practical constraint imposed by hardware/software limitations where certain IoT nodes may not be able to host certain types of ML models. Also, we build on our previous studies in the areas of distributed processing [9]–[12], green data center [13]–[22] and core networks [23]–[28], service embedding and virtualization in IoT [29]–[32], machine learning and health care systems [33]-[36] and network coding for core networks [33]–[38].

The remainder of this paper is organized as follows: Section II describes the proposed cloud fog architecture and the MILP optimization model for the distributed placement of VMs representing DNN layers. Section III provides the results and discussion while Section IV provides the conclusions and future work.

## II. THE PROPOSED CLOUD FOG NETWORK

We assume that the proposed architecture shown in Figure 1 supports full virtualization at the hardware level, which implies that different VMs can be created and deleted on the fly, regardless of the heterogeneity in the specification of the hardware equipment. We also abstract DNN workloads by random virtual topologies that are comprised of several VMs inter-connected by virtual links for data exchange. Figure 1 shows a cloud fog network with four processing layers, namely: IoT, Access Fog Node (AFN), Metro Fog Node (MFN), and the Cloud. In the Edge Network, we have several

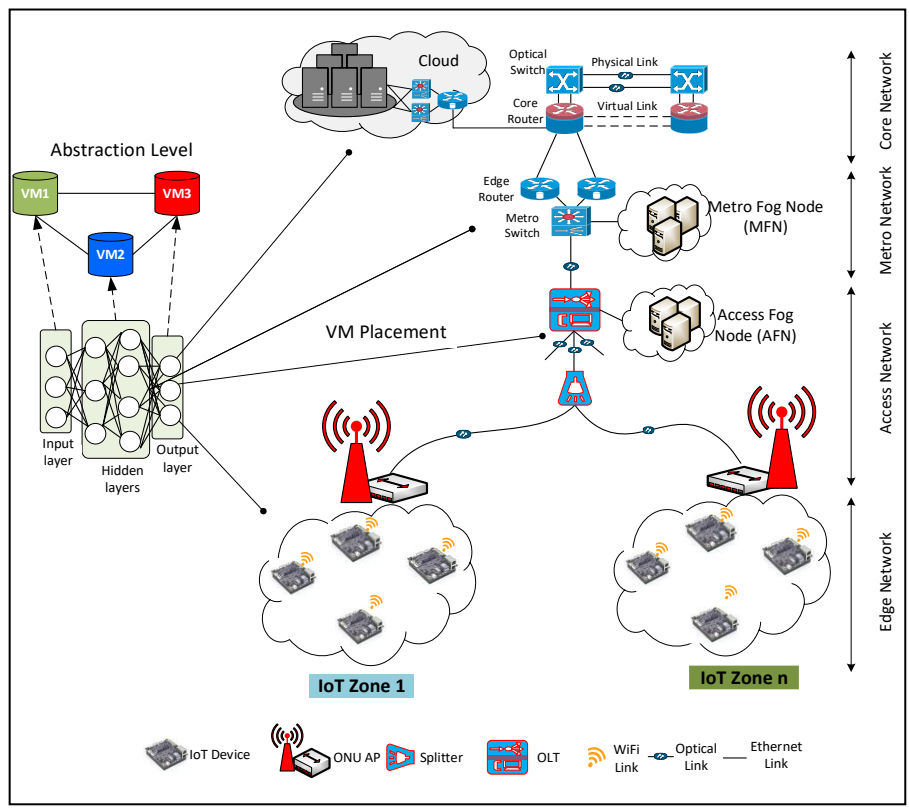

Figure 1: The Evaluated Cloud Fog over PON Access Network.

IoT devices distributed in different zones where they can collect different types of data and perform processing using their onboard CPUs. For the access network, we consider a Passive Optical Network (PON). The single PON contains several Optical Networking Units (ONUs) that connect with the IoT devices via Wi-Fi and aggregate IoT traffic through fiber links and a splitter towards an Optical Line Terminal (OLT) that can be placed at the operator's exchange office [39]. In the access layer, the Access Fog Node (AFN) containing several servers is connected to the OLT and the OLT connects to the metro network through a metro switch. The metro switch connects to the core network via multiple edge switches. The metro switch is also connected to a Metro Fog Node (MFN) which has slightly higher number of servers than the AFN due to the number of users it serves. The core network is an IP over WDM network, which is comprised of a virtual and physical layer. In the virtual layer, IP routers aggregate traffic from the access network and in the physical layer, optical switches are used to establish the connection between the core nodes [40]. In this work, we consider a single cloud data center that is one hop from the core node that aggregates traffic from the metro and access network that link the IoT nodes.

### A. MILP MODEL

The physical network shown in Figure 1 is modelled as an undirected graph $G = (N, L)$, where $N$ represents the set of all nodes and $L$ the set of links connecting those nodes in the topology. A virtual request is represented by the directed graph $G^r = (R^r, L^r)$, where $R^r$ is the set of VMs representing virtualized DNN layers and $L^r$ is the set of virtual links connecting those VMs. In this subsection, we formulate the Mixed Integer Linear Programming model (MILP) that optimizes the placement of the interconnected VMs over the proposed cloud fog architecture. Benefiting from our track record in MILP optimization and particularly in network virtualization and service embedding in [13], [14], respectively, we developed an optimization framework to optimally place virtualized DNN functions in the cloud fog network.

### B. NOTATIONS

Before introducing the optimization model, we define the sets, parameters and variables used:

**Sets**:
| | |
|---|---|
| $\mathbb{DCN}$ | Set of data center node(s). |
| $\mathbb{MFN}$ | Set of MF nodes. |
| $\mathbb{AFN}$ | Set of AF nodes. |
| $\mathbb{IoT}$ | Set of IoT devices. |
| $\mathbb{IN}$ | Set of IoT nodes generating data for the input layer. |
| $\mathbb{P}$ | Set of nodes that can process virtual requests, where $\mathbb{P} = \mathbb{DCN} \cup \mathbb{MFN} \cup \mathbb{AFN} \cup \mathbb{IoT}$. |
| $\mathbb{R}$ | Set of virtual requests. |
| $\mathbb{VM}_r$ | Set of VMs in a virtual request $r \in \mathbb{R}$. |
| $\mathbb{N}$ | Set of all nodes in the CFN architecture. |
| $\mathbb{N}_m$ | Set of neighbor nodes of node m $\in \mathbb{N}$. |

**Parameters:**
| | |
|---|---|
| $s$ and $d$ | Index the source and destination nodes of a virtual request. |
| $b$ and $e$ | Index source and destination of processing nodes hosting VM(s), where $b, e \in P, b \neq e$. |
| $m$ and $n$ | Index the physical links. |
| $P_s^r$ | $P_s^r = 1$, if in virtual request $r \in \mathbb{R}$, virtual machine $s \in VM_r$ is an input layer, otherwise $P_s^r = 0$. |
| $\pi_n^{(net)}$ | Idle power consumption of network node $n \in \mathbb{N}$. |
| $\epsilon_n$ | Energy per bit of network node $n \in \mathbb{N}$, in W/Gb/s. |
| $\pi_p^{(pr)}$ | Idle power consumption of a single CPU at node $p \in \mathbb{P}$. |
| $NS_p$ | Maximum number of CPUs deployed at processing node $p \in \mathbb{P}$. |
| $E_p$ | Energy per GFLOP of processing node $p \in \mathbb{P}$. |
| $\delta_n$ | Idle power proportion factor attributed to our IoT application on high-capacity networking equipment $n \in N$. |
| $k$ | The number of VMs that can be allocated to a single IoT node. |

**Variables:**

| | |
|---|---|
| $\lambda^{b,e}$ | Traffic demand between processing node pair $(b,e) \in \mathbb{P}$ aggregated after all VSRs are embedded. |
| $\lambda_{m,n}^{b,e}$ | Traffic demand between processing node pair $(b,e) \in \mathbb{P}$ aggregated after all VSRs are embedded, traversing physical link $(m,n)$, $m \in \mathbb{N}$ and $n \in \mathbb{N}_m$. |
| $\lambda_n$ | Amount of traffic aggregated by network node $n \in \mathbb{N}$, where $\lambda_n = \sum_{b \in \mathbb{P}} \sum_{e \in \mathbb{P}: b \neq e} \sum_{m \in \mathbb{N}} \sum_{n \in \mathbb{N}_m} \lambda_{m,n}^{b,e} + \sum_{b \in \mathbb{P}} \sum_{e \in \mathbb{P}: b \neq e} \sum_{m \in \mathbb{N}: m \neq e} \sum_{n \in \mathbb{N}_m} \lambda_{n,m}^{b,e}$. |
| $\beta_n$ | $\beta_n = 1$, if network node $n \in \mathbb{N}$ is activated, otherwise $\beta_n = 0$. |
| $\theta_p$ | Amount of traffic aggregated by processing node $p \in \mathbb{P}$. |
| $\Omega_p$ | Amount of workload in FLOPS, allocated to processing node $p \in \mathbb{P}$. |
| $N_p$ | Number of activated processing servers at processing node $p \in \mathbb{P}$. |
| $\Phi_p$ | $\Phi_p = 1$, if processing node $p \in \mathbb{P}$ is activated, otherwise $\Phi_p = 0$. |
| $\delta_b^{r,s}$ | $\delta_b^{r,s} = 1$, if virtual machine $s \in VM_r$ is embedded for processing at node $b \in P$, otherwise $\delta_b^{r,s} = 1$. |

Total power consumption is the sum of two parts: 1) network power consumption, 2) processing power consumption. The processing power consumption here includes switches routers and power consumed from servers within those nodes to provide (LAN) local area network.

- *Network Power Consumption (net_pc):*

This is given by:
$$\sum_{n \in \mathbb{N}} \epsilon_n . \lambda_n + \sum_{n \in \mathbb{N}} \beta_n . \pi_n^{(net)} . \delta_n \quad (1)$$

The power consumption of the networking equipment comprises of power consumption of routers and switches of all the nodes in the cloud fog architecture shown in Figure 1. The first term calculates the proportional power consumption (negligible in this work) and the second term accounts for the idle power consumption of the networking equipment.

- *Processing Power Consumption (proc_pc):*

This is given by:
$$\sum_{p \in \mathbb{P}} E_p . \Omega_p + \sum_{p \in \mathbb{P}} N_p . \pi_p^{(pr)} + \sum_{p \in \mathbb{P}} EL_p . \theta_p \quad (2)$$
$$+ \sum_{p \in \mathbb{P}} \Phi_p . \pi_p^{(LAN)} . \delta_n$$

Similarly, the first two terms are the proportional and idle power consumptions of processing servers, respectively. Whilst the third and fourth terms account for the proportional and idle power consumption of the LAN equipment inside server nodes.

The objective of the model is as follows:

**Minimize:** *net_pc + proc_pc*

**Subject to:**

$$\sum_{b \in \mathbb{P}} \delta_b^{r,s} = 1 \qquad \forall r \in \mathbb{R}, s \in \mathbb{VM}_r: P_s^r \neq 1 \quad (3)$$

Constraint (3) ensures that VMs of all virtual requests are embedded, except the input VMs that must be embedded onto the preselected IoT acting as the source node.

$$\sum_{s \in \mathbb{VM}_r} \sum_{b \in \mathbb{IN}} \delta_b^{r,s} = 1 \qquad \forall r \in \mathbb{R}: P_s^r = 1 \quad (4)$$

Constraint (4) ensures that input layers of virtual requests are embedded on those IoT nodes acting as source.

$$\sum_{r \in \mathbb{R}} \sum_{s \in \mathbb{VM}_r} \delta_b^{r,s} \leq k \qquad \forall b \in \mathbb{IoT}: b \notin \mathbb{II} \quad (5)$$

Constraint (5) restricts the sum of VMs allocated to a single IoT node to be less than or equal to the parameter $k$.

$$\sum_{n \in \mathbb{N}_m} \lambda_{m,n}^{b,e} - \sum_{n \in \mathbb{N}_m} \lambda_{n,m}^{b,e} = \begin{cases} \lambda^{b,e} & m = s \\ -\lambda^{b,e} & m = d \\ 0 & otherwise \end{cases} \quad (6)$$
$$\forall b, e \in \mathbb{P}, d \in \mathbb{P}, m \in \mathbb{N}: b \neq e.$$

Constraint (4) is the traffic flow conservation constraint that preserves the flow of traffic.

In this paper, due to space limitations, we have only included the key parameters, variables, and constraints. The remaining constraints deal with traffic and CPU demand realization, binary indicators and capacity constraints.

III. RESULTS AND DISCUSSIONS

In our evaluations, we used the parameters in Table 1 and Table 2 for the processing and networking devices, respectively. It is noteworthy that, where possible, device specifications were collected from manufacturer datasheets, however, we have also made simple but plausible assumptions. For instance, high-capacity networking equipment at the access, metro and core network are highly shared by many applications and services. Thus, we have assumed that only a portion of the idle power which we refer to as idle power proportion ($\delta$) is attributed to our application. The idle power consumption of core network devices is assumed to be 90% of the device's peak power consumption. As for lightly shared devices such as ONU APs, we assume the idle power is 60% of the device's peak power consumption. We also assume the cloud data center is a single hop from the aggregating core router and based on the topology of the NSFNET, the average distance between the core nodes is 509 km [41]. In total, there are 30 IoT devices, uniformly distributed into four IoT zones: IoT Zone 1 to IoT Zone 4. In each zone, there are five IoT devices that are connected through Wi-Fi links to their corresponding ONU AP. In total, we have considered four ONU APs and a single OLT device. As for the workloads, we assume that the virtual requests are issued by the IoT devices. In this work a single IoT device acts as the source node (i.e., the node that provides data for the input layer). We consider 15 virtual requests that are all embedded simultaneously on the cloud fog network. The number of virtual nodes (or VMs) per request is assumed to be between 4 – 5 nodes. The CPU demand per virtual node is randomly distributed between 0.6 – 10 GFLOPS. We assume that only the input layer nodes require negligible CPU workload as most of the intensive tasks are performed by the hidden layers. We have considered a linear power consumption profile in the MILP model; hence the proportional power consumption of networking is negligible compared to the processing power consumption. The proportional power consumption is a function of the workload whilst the idle part is consumed as soon as the device is activated. We have assumed that there are enough CPU resources at all layers to host all the workloads. Finally, the MILP model is solved using IBM's commercial solver CPLEX over the University of Leeds high performance computing facilities (ARC3) using 24 cores with 126 GB of RAM.

*A. Single VM Allocation at IoT*

We evaluated the impact of the constraint that only permits single VMs to be processed by any IoT device at a given time. Our aim was to represent a scenario in which, due to hardware/ software limitations, low power IoT nodes are not always capable of processing multiple types of VMs. Figure 2 shows the total power consumption which is the sum of the networking and the processing power consumptions against different values of the $\delta$ factor. It can be observed that when the $\delta$ factor is low (3%), the model favors the IoT and cloud layers for processing the virtual requests due to the processing efficiency of the cloud and the low power consumption of the IoT nodes. However, as shown in Figure 3, for cases where the $\delta$ is high (6% and 10%), it can be observed that the cloud is no longer a favorable choice as it loses its merit due to the power consumption of the transport network. This assumes that the idle power of the highly shared networking equipment grows linearly with the growth in the IoT applications demands. Interestingly, despite the ONU power consumption overhead, the IoT is always the predominant layer in all cases if there is enough processing capacity to host all the workloads.

TABLE I. CPU INPUT PARAMETERS

| Devices | Max(W) | Idle(W) | GFLOPS | Efficiency (W/GFLOPS) |
|---|---|---|---|---|
| IoT CPU | 7.3 [42] | 2.56 [42] | 13.5 [42] | 0.35 |
| AFN CPU | 37.2 [42] | 13.8 [42] | 34.5 [42] | 0.67 |
| MFN CPU | 37.2 [42] | 13.8[42] | 34.5 [42] | 0.67 |
| Cloud CPU | 298 [42] | 58.7 [42] | 428 [42] | 0.55 |

TABLE II. NETWORKING INPUT PARAMETERS

| Devices | Max (W) | Idle (W) | Bitrate (Gbps) | Efficiency (W/Gbps) |
|---|---|---|---|---|
| ONU Wi-Fi AP | 15 [42] | 9 [42] | 10 [42] | 0.6 |
| OLT | 1940 [42] | 60 [42] | 8600 [42] | 0.22 |
| Metro Router Port | 30 [42] | 27 [42] | 40 | 0.08 |
| Metro Switch | 470 [42] | 423 [42] | 600 | 0.08 |
| IP/WDM Node | 878 [42] | 790 [42] | 40 | 0.14 |

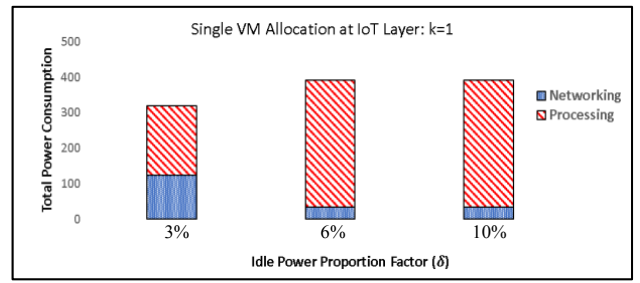

Figure 2: Total power consumption under different values of $\delta$ when k=1.

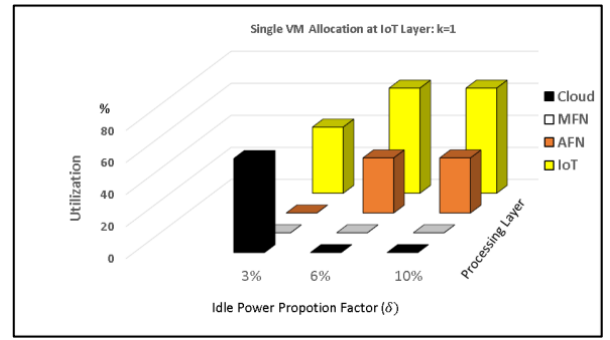

Figure 3: Workload distribution under different values of $\delta$ when k=1.

*B. Multiple VM Allocation at IoT*

In this scenario, we increase the value of the paremer *k* and thus allow multiple VMs to be processed at the IoT layer. In Figure 4 and Figure 6, the results show that flexibility in the VM allocation scheme substantially increases the power savings under all values of $\delta$ by up to 65% compared to the single allocation scenario (*k*=1). In Figure 5, the trends show that the value of $\delta$ significantly influences the choice to process VMs at the cloud because of the number of hops it takes to get there, hence lower fog layers that are close to the source are perfered, despite their processing inefficiency. As can be obserevd in Figure 5, when the VM allocation constraint is relaxed, the only layer that is utilized for processing is the IoT layer and this only changes if the assigned workload exceeds the IoT layer capacity.

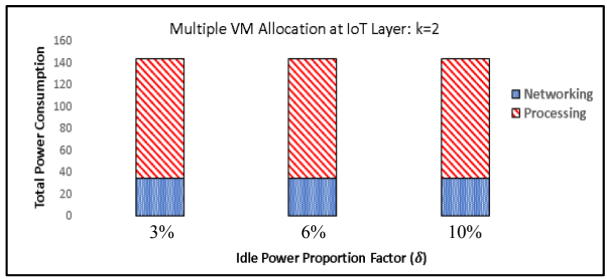

Figure 4: Total power consumption under different values of δ when k=2.

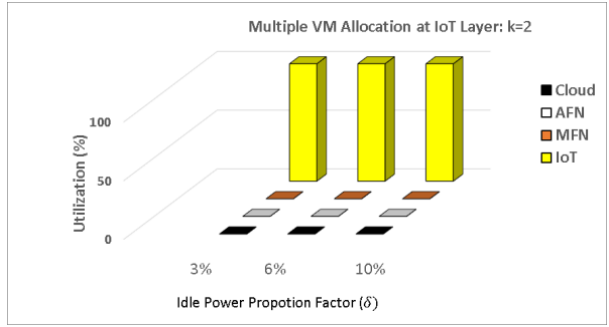

Figure 5: Workload distribution under different values of δ when k=2.

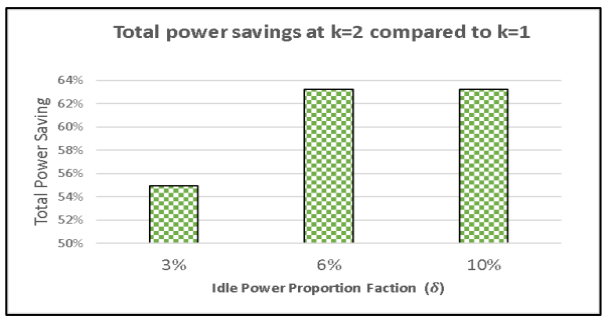

Figure 6: Total power savings achieved with k=2 compared to k=1 for different values of δ.

## IV. CONCLUSIONS

This paper extended the work in previous contributions by evaluating the impact of VM allocation flexibly against the idle power proportion factor attributed to the considered IoT application in a cloud fog architecture. The results showed substantial savings in the multiple VM allocation scenario compared to the single allocation case due to hosting all the workloads on the available IoT nodes. Future work includes the design of heuristic algorithms that can solve the problem faster compared to the MILP model and evaluating the impact of the data rate between the connected VMs on the performance of the distributed placement of VMs.

## ACKNOWLEDGMENT

The first author would like to thank the University of Tabuk for funding his PhD scholarship. The authors would like to acknowledge funding from the Engineering and Physical Sciences Research Council (EPSRC), INTERNET (EP/H040536/1), STAR (EP/K016873/1) and TOWS (EP/S016570/1) projects. All data are provided in full in the results section of this paper.